\documentclass[a4paper]{article}

\usepackage{INTERSPEECH2021}

\usepackage{pgfplots}
\DeclareUnicodeCharacter{2212}{-} 
\usepgfplotslibrary{groupplots,dateplot}
\usetikzlibrary{patterns,shapes.arrows}
\pgfplotsset{compat=newest}

\usepackage{subfigure}
\usepackage{wrapfig}
\usepackage{multirow}
\usepackage{amssymb}
\usepackage{amsmath}
\usepackage{amsthm}
\usepackage{bm}
\usepackage{xfrac}
\usepackage{nicefrac}
\usepackage{enumitem,kantlipsum}
\usepackage{algorithm}
\usepackage{algpseudocode}
\usepackage{verbatim}
\usepackage{todonotes}
\usepackage{xurl}
\usepackage{slantsc}

\algtext*{EndFor}% Remove "end for" text in algorithms
\algtext*{EndIf}% Remove "end if" text in algorithms

\usepackage{verbatim}
\usepackage{mathtools}
\usepackage{subfiles}
\usepackage{enumitem}
\usepackage[mathscr]{eucal}
\usepackage{enumitem,kantlipsum}
\usepackage{xspace}

\usepackage[most]{tcolorbox}

\PassOptionsToPackage{capitalise,noabbrev}{cleveref}
\usepackage{hyperref}
\hypersetup{
	colorlinks=true,
	linkcolor=blue,
	urlcolor=blue,
}
\usepackage{cleveref}

\newcommand{\fadam}{\textsc{FedAdam}\xspace}
\newcommand{\fedavg}{\textsc{FedAvg}\xspace}
\newcommand{\fedavgm}{\textsc{FedAvgM}\xspace}
\newcommand{\fedopt}{\textsc{FedOpt}\xspace}

\newcommand{\serveropt}{\textsc{ServerOpt}\xspace}
\newcommand{\clientopt}{\textsc{ClientOpt}\xspace}

\newcommand{\pslbl}{\textsc{PseudoLabel}\xspace}

\newcommand{\supl}{\textsc{SL}\xspace}
\newcommand{\sfl}{\textsc{SFL}\xspace}
\newcommand{\nst}{\textsc{NST}\xspace}
\newcommand{\fnst}{\textsc{FedNST}\xspace}

\title{FedNST: Federated Noisy Student Training for Automatic Speech Recognition}

\name{Haaris Mehmood, Agnieszka Dobrowolska, Karthikeyan Saravanan, Mete Ozay}

\address{
	Samsung Research UK}
\email{}
\begin{document}
	
	\maketitle
	
	\begin{abstract}
		Federated Learning (FL) enables training state-of-the-art Automatic Speech Recognition (ASR) models on user devices (clients) in distributed systems, hence preventing transmission of raw user data to a central server. A key challenge facing practical adoption of FL for ASR is obtaining ground-truth labels on the clients. Existing approaches rely on clients to manually transcribe their speech, which is impractical for obtaining large training corpora. A promising alternative is using semi-/self-supervised learning approaches to leverage unlabelled user data. To this end, we propose \fnst, a novel method for training distributed ASR models using private and unlabelled user data. We explore various facets of \fnst, such as training models with different proportions of labelled and unlabelled data, and evaluate the proposed approach on 1173 simulated clients. Evaluating \fnst on LibriSpeech, where 960 hours of speech data is split equally into server (labelled) and client (unlabelled) data, showed a \textbf{22.5\% relative word error rate reduction} (WERR) over a supervised baseline trained only on  server data.
		
		% TODO: change improvement -> word error rate reduction (WERR)
		
		%, and achieves less than X\% difference from oracle (where 100\% of data is centrally available and labelled).
		
		% >>> TODO: another number in the abstract
		
	\end{abstract}

	\section{Introduction}
	Significant improvements in Automatic Speech Recognition (ASR) have been achieved through the development of End-to-End (E2E) Attention-based models~\cite{DBLP:conf/interspeech/GulatiQCPZYHWZW20,DBLP:journals/corr/abs-2010-10504} and  semi/self-supervised learning~\cite{DBLP:journals/ml/EngelenH20,DBLP:conf/interspeech/ParkZJHCLWL20/nst-asr}, allowing for utilization of ever-increasing training corpora.
	Acquiring speech data for applications such as voice assistants requires transferring sensitive user data to the cloud, leading to privacy compromises \cite{DBLP:journals/csl/SchullerSBNVBSWEBMW15,DBLP:conf/ai/GuLCZM17,DBLP:conf/icpr/KottiK08,DBLP:conf/interspeech/SrivastavaBTV19}. 
	Federated Learning (FL)~\cite{DBLP:journals/corr/KonecnyMYRSB16,DBLP:conf/aistats/McMahanMRHA17} offers a solution by training models on user devices (clients) without sharing private data with the server. Briefly, an FL algorithm involves: (1) selecting a group of clients, (2) transmitting a \textit{global} model  to clients, (3) training the global model on local user data, (4) transmitting \textit{ gradients/weights} back to the server, (5) aggregating \textit{gradients/weights}, and (6) repeating steps 1-5 until convergence.  
	
	Various Federated ASR methods have been proposed to train ASR models in FL systems ~\cite{Dimitriadis2020AFA,DBLP:conf/icassp/GaoPZFGBL22,DBLP:conf/icassp/GulianiBM21, DBLP:conf/icassp/CuiLK21}. Specific challenges arising from data heterogeneity (speech characteristics, amount of data, acoustic environments etc.) are addressed via client-dependent data transformations \cite{DBLP:conf/icassp/CuiLK21} and imposing upper limits on the number of client samples \cite{DBLP:conf/icassp/GulianiBM21}. Improvements to distributed optimization of models, such as alternative aggregation weighting schemes based on Word Error Rate (WER)~\cite{DBLP:conf/icassp/GaoPZFGBL22} and hierarchical gradients~\cite{Dimitriadis2020AFA} have also been proposed. A realistic setup for Federated ASR is presented in \cite{DBLP:conf/icassp/GaoPZFGBL22}, showing feasibility with the French and Italian CommonVoice  subsets \cite{DBLP:conf/lrec/ArdilaBDKMHMSTW20}, comprising of thousands of challenging speakers. The above approaches all assume availability of labelled data on clients participating in FL. In real-world ASR applications, however, manual  annotation of  user data on the clients is infeasible.
	%In a real-world ASR use-cases, e.g., where a client is a single user's personal smartphone, external annotation or extensive user input are both unattainable.
	
	% on collecting, annotating and processing data from clients which may be infeasible in real-world ASR use-cases where a client is a single user's personal smartphone device.

	Recent semi- and self-supervised central training methods achieve state-of-the-art (SotA) accuracy, but require storing data at a central location \cite{DBLP:conf/nips/BaevskiZMA20,DBLP:journals/corr/abs-2010-10504,DBLP:conf/asru/ChungZHCQPW21}. A natural question then arises: can we adopt vanilla methods to leverage \textit{unlabelled} user data for federated training of ASR models? 
	
	% can we adopt vanilla methods to leverage ulabelled data for ASR in FL systems i.e. Federated ASR? 

	Semi-/self-supervised learning methods have been explored in FL systems for image and audio classification tasks \cite{DBLP:conf/icassp/Kahn0H20/selftraining/pslbl, DBLP:conf/iclr/JeongYYH21/fedsemisup, DBLP:journals/corr/abs-2108-09412,DBLP:journals/corr/abs-2110-13388, zhuang2022divergenceaware/fedselfsup}. However, these methods cannot be directly applied to tackle the above problem, since the proposed objectives are not applicable for sequence-to-sequence learning.%, or only train selected components of the model.
	%allow for only training selected components of the model on-device.
	
	A semi-supervised FL method proposed for ASR \cite{DBLP:conf/icassp/NanduryMW21} involves uploading unlabelled speech data from clients to cloud storage. An ASR model is then trained on this data using a federation of a few model trainers with high computational resources. The proposed approach is effective in a \textit{cross-silo} setup, but it is not applicable to a \textit{cross-device} setup -- a federation of thousands of user devices with low computational resources.
	
	To this end, we propose a new method called Federated Noisy Student Training (\fnst), leveraging unlabelled speech data from clients to improve ASR models by adapting Noisy Student Training (\nst)~\cite{DBLP:conf/cvpr/XieLHL20} for FL. Our work explores a challenging scenario: each client  holds and trains a model exclusively on its own unlabelled speech data, leading to a  heterogeneous data distribution,  and, more than a thousand clients participate in FL, resulting in a cross-device scenario. 
	
	% our approach performs pseudo-labelling on user devices and does not send any private user data to the cloud.
	
	%
	%the authors use unlabelled data from some clients to improve a global model.
	% for all clients. 
	% On-device SelfSup training of audio representations was explored in \cite{DBLP:journals/corr/abs-1905-11796} but without using FL. 
	
	% Methods proposed in~\cite{DBLP:journals/corr/abs-2108-09412,DBLP:journals/corr/abs-2110-13388} discuss SemiSup FL for image/vision tasks. In \cite{DBLP:conf/icassp/NanduryMW21}, ASR models are trained in a largely homogeneous setup (cross-silo FL), where unlabelled data is transferred from clients to  central servers (SemiSup FL). In \cite{DBLP:journals/corr/abs-2107-06877/fedstar}, models are trained after pseudo-labelling \cite{DBLP:conf/icassp/Kahn0H20/selftraining/pslbl} unlabelled data from clients for audio classification (SemiSup FL).

	% >>>>>>>>>  this is listed as first contribution - we are repeating ourselves?
	% To the best of our knowledge this is the first work which leverages unlabelled user data to training  ASR models in a cross-device Federated Learning scenario. 
	
	\vspace{+0.125cm}
	The contributions of this work are as follows:
	\begin{itemize}[leftmargin=*]
		\item 
		% To our best knowledge, this is the first work addressing training of ASR models distributed among clients which leverages their local private unlabelled data to improve accuracy of models in FL systems.
		To our best knowledge, this is the first work which aims to leverage private unlabelled speech data distributed amongst thousands of clients to improve accuracy of end-to-end ASR models  in FL systems.
		%proposing semi-supervised cross-device Federated Learning to improve accuracy of end-to-end  ASR models using unlabelled speech samples from clients. 
		For this purpose, we propose a new 
		%Federated ASR 
		method called \fnst, employing noisy student training for federated ASR models, which achieves 
		\textbf{22.5\% WERR over training with only labelled data.}
		% do we need to mention cross-device here?
		
		% \item We propose approaches (what approach) for limiting error increase when moving from supervised to semi-supervised training of ASR models using FedNST. Our analyses show that, 
		
		\item We elucidate the change in WER of ASR models from  central training to cross-device Federated Learning regimes with \fnst, achieving a \textbf{marginal 2.2\% relative difference from fully-centralized \nst} in a comparable setup.
		% \item We empirically analyze the importance of learning rate decay for improved convergence of Federated ASR models (4.3\% relative WERR), which has not been studied in Federated ASR literature.
		
		% To control degradation of WER, we freeze only running statistics of batch norm layers and employing per-round LR decay for Federated ASR methods.
		
		% we propose .. Our results show that ...
		
		%\item To our best knowledge, this is the first work to demonstrate benefits of using LR decay when training Federated ASR models.
		% \item We find that \fnst performed using just the pseudo-labeled data from clients attains similar level of accuracy as combining both labelled data on server and pseudo-labelled data from clients. %having original labelled data on the server is not important when training in an FL environment using a pre-trained mdoel.
	\end{itemize}

	% This paper contains the following sections, Section\ref{sec:background} discusses, .. 
	
	% \vspace{-0.25cm}
	\section{Background}
	\label{sec:background}
	%  alt title semi-sup Federated Learning
	\textbf{End-to-End (E2E) ASR} can be viewed as translating a sequence of input audio frames $\mathbf{x}= (x_1, \ldots, x_T)$, into a sequence of corresponding labels $\mathbf{y} = (y_1,\ldots,y_L)$.
	
	Modern E2E ASR models consist of an encoder-decoder architecture, trained using a weighted sum of the  sequence-to-sequence (Seq2Seq) \cite{DBLP:conf/icassp/BahdanauCSBB16} objective $\mathcal{L}_{\text{Seq2Seq}}$ and the Connectionist Temporal Classification (CTC) ~\cite{DBLP:conf/icml/GravesJ14} objective $\mathcal{L}_{\text{CTC}}$:
	\vspace{0cm}
	\begin{equation}
		\label{eq:asr-loss}
		\mathcal{L} = \nu \mathcal{L}_{\text{CTC}} + (1 - \nu) \mathcal{L}_{\text{Seq2Seq}}  
		\vspace{0cm}
	\end{equation}
	where $\nu \in [0,1]$. While the CTC objective helps with convergence during the early stages of training and is more robust to noisy conditions, the attention-based Seq2Seq objective helps in understanding long-range dependencies~\cite{DBLP:conf/icassp/KimHW17}. Recent advancements in E2E ASR models introduce self-attention to better capture intra-sequence relationships~\cite{DBLP:conf/interspeech/GulatiQCPZYHWZW20}.

	\textbf{Noisy Student Training (NST) for ASR}, a semi-supervised learning algorithm originally proposed for image classification~\cite{DBLP:conf/cvpr/XieLHL20}, has been recently shown to significantly improve ASR performance \cite{DBLP:conf/interspeech/ParkZJHCLWL20/nst-asr}. Briefly, \nst~\cite{DBLP:conf/interspeech/ParkZJHCLWL20/nst-asr} involves: (1) training an initial model $\theta_0$ on a labelled dataset $\mathcal{D}_L$; (2) integrating $\theta_0$ with a language model; (3) generating pseudo-labels for an unlabelled dataset $\mathcal{D}_U$ with $\theta_0$ (here data filtering and balancing may be applied); (4) training $\theta_0$ on a mix of $\mathcal{D}_L$ and $\mathcal{D}_U$ to generate $\theta_1$; (5) repeating steps 1-4 until convergence. 
	% \vspace{-0.0cm}
	
	\section{Proposed Approach}
	
	%In this section we describe the adaptation of \nst for Federated ASR.
	Federated Noisy Student Training (\fnst) considers a scenario where a corpus of labeled data $\mathcal{D}_L$ %${\mathcal{D}_L = \{(\mathbf{x}_n, \mathbf{y}_n) \}_{n=1}^{\vert\mathcal{D}_L\vert}}$ 
	is available on a central server. Each client $i \in \mathcal{C}$, where $\mathcal{C}$ is the set of all clients, has an unlabelled speech dataset $\mathcal{D}^i_U$. % $\mathcal{D}_U = \{\mathbf{x}^i_n, \}_{n=1}^{\vert\mathcal{D}^i_U\vert}$.
	The aim of \fnst is then to leverage the unlabelled  datasets $\mathcal{D}^i_U, \forall i \in \mathcal{C},$ to improve accuracy of an ASR model trained on $\mathcal{D}_L$ while preserving user privacy, i.e.,  without sending user data to  the server. 
	% We begin by defining the dataset conventions used in this work.
	% %Standard supervised training of an ASR model from scratch typically requires paired audios and accompanying transcripts. 
	%
	%To achieve this, we propose \fnst, an adaptation of \nst~\cite{DBLP:conf/interspeech/ParkZJHCLWL20/nst-asr} for Federated ASR . 
	%
	%
	%Below, we present simplified version of \fnst, adapting the algorithm introduced in~\cite{DBLP:conf/interspeech/ParkZJHCLWL20/nst-asr} FL:
	
	% \begin{enumerate}[noitemsep,nolistsep,leftmargin=*]
	%     \item Train an initial model, $\theta_0$, on  $D_L$ at a server. Set $\theta = \theta_0$.
	%     \item For each $i$, generate a labelled dataset, $\mathcal{D}^{i}_{\hat{U}}$ using $\pslbl$.
	%     \item Train a new local model $\theta_{i}^{t}$, $\forall i \in \mathcal{C}$.
	%     \item Send the client models to the central server.
	%     \item Aggregate $\theta_0$ and $\theta_{i}^{\dagger}$, $\forall i$ to create a new global model $\theta'$.
	%     \item Set $\theta=\theta'$ and go to 2.
	% \end{enumerate}
	%
	%
	
	% \newlength{\oldtextfloatsep}\setlength{\oldtextfloatsep}{\textfloatsep}
	\setlength{\textfloatsep}{14pt}
	{
		\small
		\begin{algorithm}[t]
			\caption{$\pmb{\fnst}$ \\ $S$ indicates a server variable and $C$ indicates a client variable.}
			\label{alg:fedopt_batched}
			\begin{algorithmic}
				\State Input: $\theta_0$, $\clientopt_{C,S}$, $\serveropt$, $\eta_{C,S}$, $\eta$, $T$, $E$
				\For{each client $i \in \mathcal{C}$ \textbf{in parallel}}
				\State $\pslbl(\theta_0)$ \Comment{generates $\mathcal{D}^{i}_{\hat{U}}$} 
				\EndFor
				\For{$t = 0, \ldots, T-1$}
				\State $\eta^{t}_{C} = \eta^{t-1}_{C} \mu^{t/\lambda}$ \Comment{Init. $\eta^{-1}_{C} = \eta_C$}
				\State Sample a subset $\mathcal{S}_t \subseteq \mathcal{C}$ of clients
				\For{each client $i \in \mathcal{S}_t$ \textbf{in parallel}}
				\State $\theta^t_{i} = \theta_{t}, \eta^{t}_{i} = \eta^{t}_{C}$ \Comment{receive $\theta_t, \eta^{t}_{C}$ from server}
				\State Retrieve: $\mathcal{D}^{i}_{\hat{U}}$ 
				\For {$e = 0, \dots, E - 1$}
				\For {$b \in \mathcal{B}_i \sim \mathcal{D}^{i}_{\hat{U}}$}
				\State $g_i = \nabla \mathcal{L}_i(\theta_i^t; b)$
				\State $\theta_{i}^t = \clientopt_C(\theta_{i}^t, g_i, \eta_i^t, e\vert\mathcal{B}_i\vert + b)$
				\EndFor
				\EndFor
				\State $\Delta_i^t = \theta^t_{i} - \theta_{t}$, $n_i = \vert\mathcal{D}^{i}_{\hat{U}}\vert$ \Comment{send $\Delta_i^t$, $n_i$ to server}
				\EndFor
				\State $n = \sum_{i \in \mathcal{S}_t} n_i$
				\State $\Delta^t_C = \sum_{i \in \mathcal{S}} \frac{n_i}{n}\Delta_i^t$
				\State $\theta^t_S = \theta_{t}$
				\For {$b \in \mathcal{B}_S \sim \mathcal{D}_L$}
				\State $g_S = \nabla \mathcal{L}_S(\theta^t_{S}; b)$
				\State $\theta^t_S = \clientopt_S(\theta^t_S, g_S, \eta_S, b)$
				\EndFor
				\State $\Delta^t_S = \theta^t_{S} - \theta_{t}$
				\State  $\Delta_t = \alpha\Delta^t_S + (1-\alpha)\Delta^t_C$
				\State $\theta_{t+1} = \serveropt(x_{t}, -\Delta_t, \eta, t)$
				\EndFor
			\end{algorithmic}
		\end{algorithm}
	}

	\subsection{Federated Noisy Student Training}
	
	% \global\setlength{\textfloatsep}{\oldtextfloatsep}
	\vspace{-0.1cm}
	Algorithm \ref{alg:fedopt_batched} describes training models using \fnst which requires an initial ASR model trained using some labelled data. This requirement is supported by several studies~\cite{DBLP:conf/icassp/RosenbergASRP17,DBLP:conf/naacl/BansalKLLG19,DBLP:conf/icassp/GaoPZFGBL22} which discuss the complexity of training modern SotA architectures from scratch with FL. 
	
	Thus, following the standard training procedure described in \cref{sec:background}, a baseline ASR model is first trained on a central server using $\mathcal{D}_L$ to produce $\theta_0$. 
	%
	% -- potentially redundant comment?
	%As $\mathcal{D}_L$ in located on a centralized server, the training data is assumed to be independent and identically distributed (i.i.d.). 
	%
	Next, the model $\theta_0$ is transferred to all clients $\mathcal{C}$. The clients use $\theta_0$ to \textit{pseudo-label} their data $\mathcal{D}^i_U$ via $\pslbl$ (Algorithm \ref{alg:pseudo-label}). A language model (LM) $\phi$ is integrated into the ASR model and a beam search~\cite{DBLP:conf/nips/SutskeverVL14} is used to improve the quality of the transcripts. 
	
	Once all clients $C$ have pseudo-labelled their data, at each FL training round $t$, the server transfers it's latest model $\theta_t$ to a randomly sampled fraction of clients $\mathcal{S}_t \subseteq \mathcal{C}$. Next, each $\textit{client } i \in \mathcal{S}_t$ trains on their pseudo-labelled data, generating a new model, $\theta_i$ (see Algorithm~\ref{alg:fedopt_batched}). The updated models are then transferred back to the server and aggregated to generate a new global model $\theta_{t+1}$.
	This process is repeated for $T$ rounds or until convergence. The optimization procedure is described in detail in \cref{sec:fedopt} and summarized in \cref{alg:fedopt_batched}.
	%
	
	% training on the clients and aggregating on the server
	
	%Being model-agnostic, as well as incurring a minimal computational overhead, \nst is an attractive candidate for \fnst of ASR models, accommodating the wide range of computational resources available across devices.
	%
	
	%Next, we consider several training modifications in order to optimize \fnst performance and efficiency. Firstly, training an ASR model from scratch with FL is extremely challenging due to the highly heterogeneous nature of user data, % [7, 13, 14]
	%optimization challenges arising from alignment sensitivity in the early stages of FL, and computational cost of training SOTA (state-of-the-art) ASR models from scratch \cite{DBLP:conf/icassp/GaoPZFGBL22}. %[21, 22]. %n large Transformers [15, 16], Transducers [17, 18] o
	%
	%Therefore, utilizing labelled data available on the server and starting with a centrally pre-trained model presents an efficient and robust approach.
	
	%Second, sending a base model to all clients may be infeasible in a large-scale FL setting. We propose alternative strategies to alleviate this:
	%\begin{enumerate}[noitemsep,nolistsep,leftmargin=*]
	%  \item The pseudo-labels can be produced using $\theta_t$ every round, as opposed to using $\theta_0$.
	%  \item Any client connecting for the first time receives two models $\theta_0$ and $\theta_t$ from the server.
	%\end{enumerate}

	We explore design choices for \fnst which balance model performance and training cost on two fronts: when to pseudo-label and whether to re-use labelled server data.
	
	First, instead of sending the model $\theta_0$ to all clients $\mathcal{C}$ at the beginning to perform \pslbl, we propose an alternative strategy: at every round $t$, the participating clients $\mathcal{S}_t$ perform pseudo-labelling using the latest global model, $\theta_t$. 
	% \begin{enumerate}[noitemsep,nolistsep,leftmargin=*]
	%   \item Using $\theta_t$.
	%   \item Using $\theta_0$ sent by the server alongside $\theta_t$ at every round.
	% \end{enumerate}
	
	% The above two strategies can be successfully used in an environment with a large number of clients and low sampling ratio i.e. when  $\vert\mathcal{S}_t\vert < \vert\mathcal{C}\vert$ where $\vert\vert$ denotes set cardinality. 
	
	% In our experiments we train models with $\vert\mathcal{S}_t\vert/\vert\mathcal{C}\vert = 7\%$. 
	
	% >>>>>>> IMPORTANT COMMENT DO NOT DELETE
	% e.g. $\le0.01$ clients sampled per round and $\ge100k$ clients in total.
	
	% >>>>>> IMPORTANT COMMENT DO NOT DELETE
	% This is more of an empirical research question, maybe we can move it to another section
	Secondly, the standard \nst~\cite{DBLP:conf/interspeech/ParkZJHCLWL20/nst-asr} procedure described in Section~\ref{sec:background} involves \textit{mixing} samples from the labelled and pseudo-labelled datasets for each training step. However, in supervised FL, it is common to start with a model pre-trained on the server with some labelled data, and during the FL process to only use client data for model updates~\cite{Dimitriadis2020AFA, DBLP:conf/icassp/GaoPZFGBL22}. In this work, we perform an empirical study to determine  the optimal strategies for data mixing with \fnst -- only aggregating client updates, or also incorporating the supervised data on the server, which was used for pre-training $\theta_0$.
	
	%-- in terms of both model accuracy and training time.
	\vspace{-0.1cm}
	
	% \newlength{\oldtextfloatsep}\setlength{\oldtextfloatsep}{\textfloatsep}
	% \setlength{\textfloatsep}{0pt}
	{
		\small
		\begin{algorithm}[t!]
			\caption{$\pmb{\pslbl}(\theta)$ - for every client $i \in \mathcal{C}$.} 
			\label{alg:pseudo-label}
			\begin{algorithmic}
				\State Input from server: $\theta$
				\State Retrieve: $\phi$, $\mathcal{D}^i_U$
				\State Initialize: $\mathcal{D}^{i}_{\hat{U}} = \varnothing$
				\For{$j = 1, 2,\ldots, \vert \mathcal{D}^i_U \vert $}
				\State $\mathbf{\hat{y}}^i_j = f(\mathbf{x}^i_j; \theta; \phi)$
				\State $\mathcal{D}^{i}_{\hat{U}} \leftarrow (\mathbf{x}^i_j, \mathbf{\hat{y}}^i_j) $
				\EndFor
				\State Store $\mathcal{D}^{i}_{\hat{U}}$ on client $i$ for future retrieval
			\end{algorithmic}
		\end{algorithm}
	}
	
	\subsection{Federated Optimization for \fnst}
	% \global\setlength{\textfloatsep}{\oldtextfloatsep}
	\label{sec:fedopt}
	\vspace{-0.1cm}
	%Next, we describe the optimization process used to train \fnst.
	
	% Since all clients require an initial model for pseudo-labelling as shown in Algorithm \ref{alg:pseudo-label}, it is paramount that a pre-trained model is used for the FL training, this is defined as $\theta_0$.
	
	The optimization procedure described in Algorithm~\ref{alg:fedopt_batched}, adopts the notation introduced in~\cite{DBLP:conf/iclr/ReddiCZGRKKM21/FedOpt}.
	%such as `\textit{pseudo-gradients}', $\serveropt$ and $\clientopt$. 
	$\mathcal{L}(\cdot)$ denotes the ASR loss defined in \eqref{eq:asr-loss}. We define two optimizers \cite{DBLP:conf/iclr/ReddiCZGRKKM21/FedOpt}, $\clientopt$ and $\serveropt$, which are used on the clients and on the server, respectively. Each performs a single step of gradient descent based on the weights $\theta$, gradients $g$ (or pseudo-gradients $-\Delta$), learning rate $\eta$ and optimization step-count. All clients individually train on their own pseudo-labelled data using  $\clientopt_C$  without sending this data to the server. The optimizer $\clientopt_S$ is used to train on the labelled data which exists on the server. 
	
	% Both optimizers are denoted $\clientopt$ because they should follow the same underlying algorithm (e.g. SGD) for consistency but can have differing hyper-parameters (e.g. $eta$).
	
	% \vspace{-0.1cm}
	For each round $t$ in FL, the same model $\theta_t$ is updated in parallel on both the server and on a set of randomly sampled clients $\mathcal{S}_t$. The resulting \textit{weights difference} $\Delta^t_{i} = \theta_i^t - \theta_t$ is uploaded to the server by each participating client $i$ after their local training for $E$ epochs. This is aggregated by the server to give $\Delta^t_C$. Similarly, a \textit{weights difference} $\Delta^t_S$ is produced after server training. The hyper-parameter $\alpha$ performs a weighted average of $\Delta^t_C$ and $\Delta^t_S$ to produce \textit{pseudo-gradients} $-\Delta_t$. Next, $-\Delta_t$ is passed to \serveropt to produce a set of weights for the next round $\theta_{t+1}$. This process is repeated for a pre-determined number of rounds, $T$, or until convergence of models as measured via WER on a validation set located on the server. 
	
	% smaller negative sign above^

	\section{Methodology}
	
	%\subsection{Evaluation Dataset}
	%Due to the permutation invariance of weights of DNN models, it is challenging to train models distributed among clients from scratch.
	%We follow the setup proposed in \cite{Dimitriadis2020AFA} for initializing federated models. 
	\textbf{Evaluation Datasets: } We evaluate our proposed method on the LibriSpeech (LS) dataset~\cite{librispeech}. Following the setup described in \cite{Dimitriadis2020AFA}, we divide the dataset into two equal subsets, $\mathcal{A}$ and $\mathcal{B}$, each comprising 480h. The two sets are \textit{disjoint}, such that all data arising from a single speaker is assigned to only one set.
	%
	% randomly assign each speaker, as given by LS speaker ID, to one of two disjoint sets $\mathcal{A}$ and $\mathcal{B}$. Both splits are equivalent (in terms of hours) and together form 960h of labelled audio. The purpose of each split is described in Table \ref{tab:ls_splits}. For FL experiments, each speaker in $\mathcal{B}$ is regarded as a client, resulting in 1173 clients.
	% data heterogeneity / non-iid data 
	% Set $\mathcal{A}$, \textit{seed data}, is used to pre-train the initial model and may be re-used later for server training, as discussed in Algorithm~\ref{alg:fedopt_batched}. Set $\mathcal{B}$ is used for \fnst exclusively
	%  such that each set has roughly the same number of hours. 
	
	In the experimental analyses, we explore two scenarios: 
	\begin{enumerate}[noitemsep,nolistsep,leftmargin=*]
		\item Set $\mathcal{A}$ is used only for initial model pre-training, i.e., an FL step is  performed using only set $\mathcal{B}$ (no data mixing).
		\item Set $\mathcal{A}$ is mixed with  $\mathcal{B}$ during federated training of models as discussed in Algorithm~\ref{alg:fedopt_batched}.
	\end{enumerate}
	
	% The terminology is as follows:
	% \begin{itemize}
	%   \item Supervised Learning (\supl): Learning on the server with true labels
	%   \item Semi-supervised Learning (SSL): Learning on the server (centrally) with true and pseudo labels.
	%   \item Semi-supervised Federated Learning (SSFL): Learning on clients with pseudo labels, (optionally) 
	%   learning on clients with true labels and (optionally) learning on the server with true labels.
	% \end{itemize}

	% The experiment terminology is defined as follows:
	% \begin{enumerate}[label=\roman*,noitemsep,nolistsep,leftmargin=*]
	%   \item \textit{SemiSup X-Y}: X hours of labelled data and Y hours of pseudo-labelled data on server (if Y $> 0$).
	%   \item \textit{SemiSupFL X-Y-Z}: X hours of labelled data on server, Y hours of labelled clients as an aggregate (if Y $> 0$) and Z hours of pseudo-labelled clients as an aggregate (if Z $> 0$). 
	%   \item \textit{SemiSupFL X-Y:Z}: X hours of labelled data on server, Y ratio of data on each client as labelled and Z ratio of data on each client as pseudo-labelled. 
	% \end{enumerate}
	% In (i), training occurs centrally. In (ii) and (iii) models are trained on clients and optionally on server (if X $> 0$). If unlabelled hours are 0 then the training is fully supervised.
	%
	%
	% \subsection{Model Architecture}
	% \vspace{-0.1cm}
	% We use an end-to-end ASR model architecture\footnote{\scriptsize \nolinkurl{https://github.com/speechbrain/speechbrain/blob/main/recipes/LibriSpeech/ASR/transformer/hparams/conformer_small.yaml}} \cite{speechbrain-repo} with a Conformer (S) encoder  \cite{DBLP:conf/interspeech/GulatiQCPZYHWZW20}, a transformer decoder and  joint CTC/seq2seq loss. 
	%
	%
	%
	\noindent
	\textbf{Experimental Setup: } We use an end-to-end ASR model architecture %\footnote{\scriptsize \nolinkurl{https://github.com/speechbrain/speechbrain/blob/main/recipes/LibriSpeech/ASR/transformer/hparams/conformer_small.yaml}}
	\cite{speechbrain-repo} with a Conformer (S) encoder  \cite{DBLP:conf/interspeech/GulatiQCPZYHWZW20}, a transformer decoder and a joint CTC+Seq2Seq objective. 
	For \cref{alg:fedopt_batched}, we set $T = 1000$, ${E = 1}$, ${\eta_{C,S} = 0.1}$, ${\eta = 1.0}$, $\alpha = 0.5$ and $\frac{\vert\mathcal{S}_t\vert}{\vert\mathcal{C}\vert} = 7\%$. We set an equivalent number of epochs for the centralized experiments. After training, the model with the lowest  WER on the dev-clean is selected. Results reported on the dev-clean, test-clean and test-other sets of LS are obtained by fusing the ASR model with an off-the-shelf language model \cite{speechbrain-repo} and using a beam search of size 10. For the federated experiments, we use a FL simulation platform with design characteristics similar to those described in  \cite{Dimitriadis2020AFA}.

	\noindent \textbf{Batch Normalization for \fnst:}
	% In large scale FL, BN often gives rise to convergence issues due to data heterogeneity.  Several solutions have been proposed   In our  experiments, we found that BN causes loss divergence in FL. 
	As reported in other works \cite{DBLP:conf/icml/HsiehPMG20/groupnorm-fl, DBLP:conf/iclr/Diao0T21/static-bn, DBLP:conf/iclr/LiJZKD21/fedbn}, we find that using standard Batch Normalization (BN) for FL gives rise to convergence issues due to data heterogeneity.
	To address this issue, we replace all BN layers with a modified version of static Batch Normalization (sBN) \cite{DBLP:conf/iclr/Diao0T21/static-bn}. sBN does not keep track of running statistics, i.e., the moving-average mean and variance associated with BN layers, during FL training. At the end of FL training, it queries all clients sequentially to produce global BN statistics which can then be used to evaluate the trained model. This post-processing step has computational and privacy concerns \cite{DBLP:conf/iclr/Diao0T21/static-bn} and it makes it difficult to perform model evaluation during training.
	
	Our modification is to simply re-use the running statistics from $\theta_0$ -- the model trained using the supervised data corpus $\mathcal{D}_L$ located at the server. The behaviour of our modified sBN layers during FL training is the same as original BN: normalization using batch mean and variance, followed by re-scale and shift using the trainable parameters $\gamma$ and $\beta$. However, it is worth noting that this approach assumes that the client and server data arise from the same data distribution i.e. we expect running statistics before and after FL to be the same.
	
	% why we chose this approach -- because it was minimally invasive in terms of arch changes to the model or re-training the pre-trained model.
	
	% which estimates the BN parameters , as well as the running statistics, i.e., the  mean and variance of the features, using a small data subset before FL training begins, and does not update the parameters or the statistics during training. 
	% % but use them for inference
	
	% Adapting sBN for \fnst, the running statistics and parameters, which are used to initialize BN for federated training, are estimated using the supervised data corpus located at the server which is used to train the initial model $\theta_0$.
	%
	% as most SotA ASR models contain BN layers
	% For the method to 
	% and modify it to be compatible with \fnst.
	% Specifically, we make two changes:  and 
	% which requires a centrally pre-trained model. To make our technique compatible with most off-the-shelf pre-trained ASR models (which contain BN layers), 
	%we follow an approach that is least intrusive: 
	%
	%  In the analyses, we find that  best results are obtained by freezing the estimated running statistics but continually updating  $\gamma$ and $\beta$ parameters using client data. %, i.e., that their estimated local statistics are similar.
	
	% ADD BELOW LINE AS SECOND LAST LINE IF SPACE
	% It avoids the privacy concerns of post-training statistics collection \cite{DBLP:conf/iclr/Diao0T21/static-bn} and is much simpler to implement than \cite{DBLP:conf/iclr/LiJZKD21/fedbn}
	
	\noindent \textbf{Federated Optimization in FedNST:}
	Using the $\fedopt$ formulation for federated optimization \cite{DBLP:conf/iclr/ReddiCZGRKKM21/FedOpt}, we examined the use of adaptive (\fedavgm and \fadam) and standard (\fedavg ~\cite{DBLP:conf/aistats/McMahanMRHA17}) server-side optimizers for FL. We found that the differences in accuracy between these optimizers were limited (results not shown due to limited space), and hence choose the computationally more efficient  \fedavg in our study. 
	
	% Since our main scope is semi-supervised FL, we omit plots of their learning curves due to lack of space.
	
	For the client-side optimizer, our default configuration \cite{speechbrain-repo} uses Adam to train the model. Our experiments indicate that using Adam as a local optimizer performs worse than using SGD  because a cross-device FL system expects stateless clients, whereas Adam has stateful parameters. This was also found to be the case in \cite{Dimitriadis2020AFA}. Thus, we use the SGD optimizer for clients. We set $\alpha$ defined in Algorithm \ref{alg:fedopt_batched} to $0.5$ for all experiments as we found this to be the best value.
	
\begin{figure}[t]
		\centering
		\resizebox{\columnwidth}{!}{%
			\input{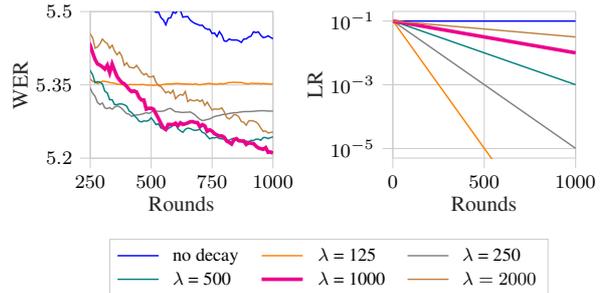}
		}
		\vspace{-0.5cm}
		\captionsetup{font={small, it}}
		\caption{ WER of models on the dev-clean  (greedy search and no LM) (left) and client LR (right) for various $\lambda$ values. Plot on the left is smoothed using exponential moving average with weight 0.8 and only the last 750 rounds are shown. Plot on the right is logarithmically scaled along the y-axis.}
		\label{fig:plot_decay_steps_expt}
		%\vspace{-0.5cm}
	\end{figure}

	\noindent \textbf{Learning Rate (LR) Decay for Clients in FedNST:}
	When data is distributed across clients (e.g.\ one client per speaker), such a dataset is considered to be non-iid. Li et al.\ \cite{DBLP:conf/iclr/LiHYWZ20/fedavg-convergence-non-iid} reported that the LR of federated optimizers must decay to guarantee convergence when training with non-iid data. Thus, we use LR decay to seek a better global minima in the analyses.
	Among various LR reduction methods \cite{DBLP:journals/corr/abs-2007-00878/lr-dec}, we choose to only decay the client LR $\eta^t_C$ at the start of each round $t$ and keep it fixed for local training. We use exponential decay as described in Algorithm \ref{alg:fedopt_batched} where the LR $\eta^t_C$ is shared amongst participating clients $\mathcal{S}_t$, and updated at each round. %Other decay approaches can also be used.
	We experimented with various values for $\mu$ (decay rate) and $\lambda$ (decay steps) defining the LR decay curve. Training curves used for selecting a suitable value of $\lambda$ are shown in Figure~\ref{fig:plot_decay_steps_expt}. We found $\lambda = 1000$ to perform the best, and used this for all subsequent experiments. Therefore, for our setup, using a weak decay is more effective than strong decay.

	%We studied a recent FL aggregation weighting scheme for ASR, negative training loss results \cite{Dimitriadis2020AFA, DBLP:conf/icassp/GaoPZFGBL22} and did not find any to be particularly better than the default weighting scheme. This could be due to the characteristics of LibriSpeech dataset \cite{Dimitriadis2020AFA}. We couldn't study the affect of WER based weighting because it is very inefficient to evaluate every sampled client's model for every round using the validation set. We did not study the affect of a recently published diversity-based scaling as it was shown to only work in a cross-silo setting with much fewer clients than us \cite{DBLP:conf/icassp/NanduryMW21}.
	
	%\subsubsection{FedProx}
	%We experimented with FedProx \cite{DBLP:conf/mlsys/LiSZSTS20/fedprox} which adds a proximal loss to aid in training with heterogeneous data. We found that lower values for the proximal weight didn't give any difference compared to FedAvg and higher values showed divergence of dev-clean evaluated WER curve after certain number of rounds. This is likely because the FedProx loss dominates after certain rounds of training since the model converges near to a global minima which is empirically the average of all client's local minimas. When this happens the loss term for unsupervised and supervised is low and the loss term for fedprox is high. To alleviate this we can make use of appropriate lr scheduling but this doesn't solve the root cause. We propose one can improve this by having a decaying fedprox term. We decided not to use fedprox for all our experiments.
	% \vspace{-0.125cm}
	
	% It is recently suggested in FL literature that once fl training converges near the region of global minima the training might either plateu or diverge \cite{DBLP:journals/corr/abs-2007-00878/lr-dec}  as the different clients near the global minima give strongly different directions of training via their pseudo-gradients this causes the overall global model training to plateu/diverge. To alleviate this there are several proposals we make use of the simplest one which is LR decay, 
	
	% \subsubsection{Per Round Pseudo-Labelling}
	% \todo[inline]{add or delete section}
	
	% \subsubsection{Experimenting with different $\alpha$}
	% \todo[inline]{add or delete section}
	
	\begin{table}[h!]
		\vspace{-0.15cm}
		\caption{Description of data splits used for experiments.}
		\vspace{-0.3cm}
		\label{tab:ls_splits}
		\centering
		\resizebox{0.85\columnwidth}{!}{%
			\begin{tabular}{lllll}
				\toprule
				\textbf{Split} & \textbf{\begin{tabular}{@{}c@{}}Hrs\end{tabular}} & \textbf{\begin{tabular}{@{}c@{}} Num. Spks\end{tabular}} & \textbf{Supervision} &\textbf{Location} \\
				\midrule
				$\mathcal{A}$       & 480        & 1165         & Labelled &  Server        \\
				$\mathcal{B}$      & 480        & 1173         & Unlabelled & Clients \\
				\bottomrule
			\end{tabular}
		}
		\vspace{-0.2cm}
	\end{table}

	\section{Experimental Analyses and Results}
	
	\subsection{Comparison with SotA Methods}
	%The main table of results are presented Others are not main?
	
	In Table \ref{tab:main-results}, we compare WER of models trained using centralized supervised learning (\supl), supervised Federated Learning (\sfl), \nst, and our proposed \fnst. 
	
	The results given in the first row (indicated by $\supl_{\text{seed}}$) are obtained through training models on $\mathcal{A}$ (labelled). 
	The results given in the second row (indicated by \supl) are obtained by training models on $\mathcal{A} \cup \mathcal{B}$ (both labelled). Here, \supl is the lower-bound or best possible WER. We present 2.59 $\%$ WER on test-clean with 960h of labelled data which is close to SotA for a comparable sized model \cite{DBLP:conf/interspeech/GulatiQCPZYHWZW20}.
	The last three rows present, \nst, \sfl and \fnst, trained over a combination of $\mathcal{A}$ and $\mathcal{B}$, starting from the pre-trained model $\supl_{\text{seed}}$. All \fnst experiments use $\supl_{seed}$ as $\theta_0$ from Table \ref{tab:main-results} unless stated otherwise. For simplicity, in our study, both \nst and \fnst models are trained for a single generation and without any data filtering or balancing methods defined in \cite{DBLP:conf/interspeech/ParkZJHCLWL20/nst-asr}.
	
	%Our results show that moving from central server trained \nst to \fnst leads to almost no degradation in WER.
	
	% When training with pseudo-labelled $\mathcal{B}$ and labelled $\mathcal{A}$
	\fnst achieves comparable WER over \nst (2.2\% relative difference for test-clean), without the need of sending user data to a central server. We further observe a relative \textit{WER increase} of 8.5\% when comparing \sfl with \fnst. This is still smaller than the 22.5\% relative \textit{WER reduction} achieved by \fnst over $\supl_{seed}$. These results provide strong motivations to use \fnst in a real-world scenario. 
	
	%\vspace{-0.125cm}
	
	\begin{table}[t]
		\captionsetup{font=small}
		\caption{
			Comparison of WER (\%) of models trained under different labelled data regimes.}
		\vspace{-0.25cm}
		\label{tab:main-results}
		\centering
		\resizebox{0.9\columnwidth}{!}{%
			\begin{tabular}{lllccc}
				\toprule
				\textbf{Method}  & L & U   & \textbf{test-clean} & \textbf{test-other} \\ 
				\midrule
				$\text{\supl}_{\text{seed}}$ & $\mathcal{A}$ & $\varnothing$      & 4.27         & 8.99 \\
				\supl & $\mathcal{A}\cup\mathcal{B}$ & $\varnothing$         & 2.59   & 6.30  \\
				\textbf{\nst}   & $\mathcal{A}$ & $\mathcal{B}$     & \textbf{3.24}         & \textbf{7.96} \\
				\midrule
				\sfl & $\mathcal{A}\cup\mathcal{B}$ & $\varnothing$  & 3.05  & 7.57 \\
				\textbf{\fnst} & $\mathcal{A}$ & $\mathcal{B}$      & \textbf{3.31}        & \textbf{8.07}   \\
				\bottomrule
			\end{tabular}
		}
		
	\end{table}
	% \vspace{-0.5cm}

	% FOR BACKUP ONLY -- SAME RESULTS
	% \begin{table}[t]
	%   \captionsetup{font=small}
	%   \caption{
	%   Comparison of WER (\%) of models trained under different labelled data regimes.}
	%   \vspace{-0.25cm}
	%   \label{tab:main-results}
	%   \centering
	%     \resizebox{\columnwidth}{!}{%
	%   \begin{tabular}{lllccc}
	%     \toprule
	%   \textbf{Method}  & L & U   & \textbf{dev-clean} & \textbf{test-clean} & \textbf{test-other} \\ 
	%     \midrule
	%     $\text{\supl}_{\text{seed}}$ & $\mathcal{A}$ & $\varnothing$    & 4.05        & 4.27         & 8.99 \\
	%     \supl & $\mathcal{A}\cup\mathcal{B}$ & $\varnothing$       & 2.45  & 2.59   & 6.30  \\
	%         \textbf{\nst}   & $\mathcal{A}$ & $\mathcal{B}$  & \textbf{3.21}        & \textbf{3.24}         & \textbf{7.96} \\
	%     \midrule
	%     \sfl & $\mathcal{A}\cup\mathcal{B}$ & $\varnothing$     & 2.87 & 3.05  & 7.57 \\
	%     \textbf{\fnst} & $\mathcal{A}$ & $\mathcal{B}$    & \textbf{3.10}        & \textbf{3.31}        & \textbf{8.07}   \\
	%     \bottomrule
	%   \end{tabular}
	%   }
	% % \vspace{-0.5cm}
	% \end{table}
	% % \vspace{-0.5cm}
	
	% In Table~\ref{tab:sec1a}-\ref{tab:sec2b}, batch size is 16 for training and a beam size is 10, batch size is 8 for evaluation and 16 for training.
	
	\begin{table}[t]
		\caption{An analysis of the relationship between the duration of labelled data  $\vert \mathcal{A} \vert$ in hours and WER (\%) of models.}
		\vspace{-0.25cm}
		\label{tab:sec3}
		\centering
		\resizebox{0.94\columnwidth}{!}{%
			\begin{tabular}{@{\extracolsep{8pt}}ccccc@{}}
				\toprule
				\multirow{2}{*}[-1.5pt]{\textbf{$\vert \mathcal{A} \vert$}}  & \multicolumn{2}{c}{\textbf{$\text{\supl}_{\text{seed}}$}} & \multicolumn{2}{c}{\textbf{\fnst}} \\
				\cmidrule{2-3} \cmidrule{4-5}
				& \textbf{test-clean} & \textbf{test-other} & \textbf{test-clean} & \textbf{test-other} \\
				\midrule
				120  & 9.44	        & 18.0         & 7.89	       & 15.9                \\
				240  & 6.96         & 12.9         & 5.30         & 12.0               \\
				360  & 5.52         & 11.0         & 4.69         & 10.0                 \\
				480  & 4.27         & 8.99          & 3.31         & 8.07 \\ % NOTE: THIS result must match with 480-0-480, ensure that is the case!
				\bottomrule
			\end{tabular}
		}
		%  \vspace{-0.35cm}
	\end{table}

	\subsection{Analyses with Different Pseudo-Labelled Data Regimes}
\begin{figure}[t]
		\centering
		\resizebox{\columnwidth}{!}{%
			% This file was created with tikzplotlib v0.10.1.
\begin{tikzpicture}

\definecolor{lightgray204}{RGB}{204,204,204}
\definecolor{darkslategray38}{RGB}{38,38,38}
\definecolor{darkgray176}{RGB}{176,176,176}
\definecolor{steelblue31119180}{RGB}{31,119,180}

\pgfplotsset{
% override style for non-boxed plots
    % which is the case for both sub-plots
    every non boxed x axis/.style={} 
}

\begin{groupplot}[
    group style={
        group name=my fancy plots,
        group size=2 by 2,
        xticklabels at=edge bottom,
        vertical sep=0pt,
        horizontal sep=1cm,
    },
    width=0.6\columnwidth,
    xmin=-10, xmax=110,
]

% sec1a,2a -- A+B, A -- changing % labelled clients as whole -- test-other
\nextgroupplot[
                ymin=6.75,ymax=9.0,
                ytick={7.5, 8.0, 8.5},
                xtick={0, 50, 100},
                axis x line=top,
                axis y discontinuity=parallel,
                height=3.5cm,
                % legend cell align={left},
                title={(a)},
                title style={yshift=-0.15cm},
                legend columns=4,
                legend style={
                  fill opacity=1,
                  draw opacity=1,
                  text opacity=1,
                  /tikz/every even column/.append style={column sep=0.25cm},
                  at={(1.125,-1.70)},
                  nodes={scale=0.9, transform shape},
                 anchor=north,
                  draw=lightgray204
                },]

\addlegendimage{no markers,blue}
\addlegendimage{no markers,red}
\addlegendimage{only marks, mark=square*, mark options={scale=1.0}}
\addlegendimage{only marks, mark=triangle*, mark options={scale=1.5, yshift=-0.25}}

\addplot[color=blue,mark=triangle*, forget plot]
table {% 1a - test-other
0 8.07
25 7.96
50 7.92
75 7.79
100 7.57
};
\addplot[color=red,mark=triangle*, forget plot]
table {% 2a - test-other
0 8.46
25 8.17
50 7.95
75 7.83
100 7.61
};

\legend{$\mathcal{A}\cup\mathcal{B}$, $\mathcal{B}$, test-clean, test-other}

% sec1b,2b -- A+B, A -- changing % data labelled per client -- test-other
\nextgroupplot[
                ymin=6.75,ymax=9.0,
                ytick={7.5, 8.0, 8.5},
                xtick={0, 50, 100},
                title={(b)},
                title style={yshift=-0.15cm},
                axis x line=top,
                axis y discontinuity=parallel,
                height=3.5cm,]
\addplot[color=blue,mark=triangle*, forget plot]
table {% 1b - test-other
0 8.07
25 8.01
50 7.86
75 7.72
100 7.57
};
\addplot[color=red,mark=triangle*, forget plot]
table {% 2b - test-other
0 8.46
25 8.17
50 8.01
75 7.79
100 7.61
};

% sec1a,2a -- A+B, A -- changing % labelled clients as whole -- test-clean
\nextgroupplot[ymin=2.9,ymax=3.6,
               ytick={3.0, 3.25, 3.5},
               axis x line=bottom,
               xtick={0, 50, 100},
               ylabel style={xshift=1.2cm, yshift=-0.2cm},
               ylabel=\textcolor{darkslategray38}{WER ($\%$)},
               xlabel style={yshift=-0.25cm},
               xlabel=\textcolor{darkslategray38}{Labelled clients ($\%$)},
               height=3.5cm,]
\addplot[color=blue,mark=square*, mark options={scale=0.75}, forget plot] %smooth,
table {% 1a - test-clean
0 3.31
25 3.25
50 3.22
75 3.11
100 3.05
};
\addplot[color=red,mark=square*,mark options={scale=0.75}, forget plot] % smooth
table {% 2a - test-clean
0 3.45
25 3.33
50 3.19
75 3.13
100 3.03
};

% sec1b,2b -- A+B, A -- changing % data labelled per client -- test-clean
\nextgroupplot[ymin=2.9,ymax=3.6,
               ytick={3.0, 3.25, 3.5},
               axis x line=bottom,
               xtick={0, 50, 100},
            %   ylabel style={xshift=1.0cm},
            %   ylabel=\textcolor{darkslategray38}{WER ($\%$)},
               xlabel style={align=center,text width=2.5cm},
               xlabel=\textcolor{darkslategray38}{Labelled samples \newline per client ($\%$)},
               height=3.5cm,]
\addplot[color=blue,mark=square*, mark options={scale=0.75}, forget plot] %smooth,
table {% 1b - test-clean
0 3.31
25 3.25
50 3.17
75 3.14
100 3.05
};
\addplot[color=red,mark=square*,mark options={scale=0.75}, forget plot] % smooth
table {% 2b - test-clean
0 3.45
25 3.31
50 3.19
75 3.11
100 3.03
};

\end{groupplot}

\end{tikzpicture}
		}
		\vspace{-0.5cm}
		\captionsetup{font={small, it}}
		\caption{Reduction of WER as the percentage of labelled clients (i) or samples per client (ii) increases. Relative WER increases from labelled (100\%) to pseudo-labelled (0\%) data regime.} %with \textbf{convergence of relative WER to 7.6\%}.}
		\label{fig:plot_changing_prop_labelled}
		%\vspace{-0.5cm}
	\end{figure}
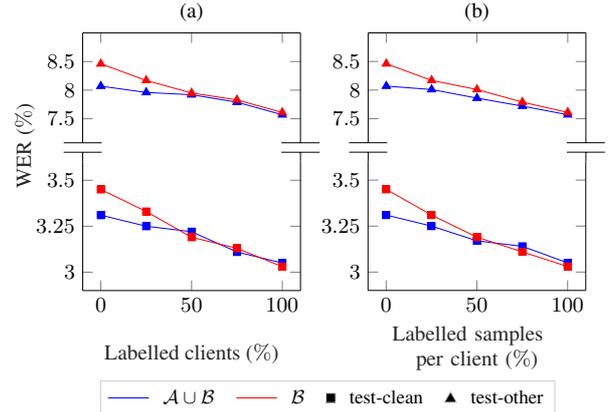

	% Generally, the benefits of not requiring manual labelling and sending user data to a central server should still outweigh this drawback.
	\vspace{-0.1cm}
	Intuitively, models trained with fully supervised data perform better than those trained with self-training methods (e.g. NST) on partially labelled data. This is due to the noise induced by inaccurately predicted labels. We explore how changing the proportion of labelled and pseudo-labelled data affects model performance. The purpose is to: (1) quantify the hypothetical gain in performance if ground-truth labels were available for all clients, and (2) empirically verify if $\mathcal{A}$ is useful during federated training. The results are depicted in \cref{fig:plot_changing_prop_labelled}. 
	
	\cref{fig:plot_changing_prop_labelled} presents two analyses: \cref{fig:plot_changing_prop_labelled}~(i) varies the percentage of clients with entirely labelled data against clients which pseudo-label their data (i.e., applying Algorithm~\ref{alg:pseudo-label}). \cref{fig:plot_changing_prop_labelled}~(ii) varies percentage of labelled samples per client, i.e., ratio of samples with ground-truth against samples that are pseudo-labelled \textit{for each client}.
	Each plot in \cref{fig:plot_changing_prop_labelled} presents two sets of graphs showing results from test-clean and test-other, respectively. Within each set, graphs representing $\mathcal{A} \cup \mathcal{B}$ (blue) and $\mathcal{B}$ (red) show the impact of dataset mixing in FL. Percentage of labelled clients and labelled samples per client is thereafter referred to as \textit{`x\%-labelled'} for brevity. 
	
	Figures~\ref{fig:plot_changing_prop_labelled}~(i) and (ii) show similar trends, since in both cases the cumulative hours of labelled vs. unlabelled data obtained from all clients are roughly the same. With reference to  (1), we find that \textit{0\%-labelled} leads to a relative WER increase of 8.5\% compared to \textit{100\%-labelled} for the $\mathcal{A} \cup \mathcal{B}$ setup. With reference to (2), training models with FedNST using only $\mathcal{B}$ does not cause significant catastrophic forgetting. Instead, the model retains its ability to further learn from new data. We find that when using \textit{0\%-labelled}, re-using $\mathcal{A}$ (i.e., data-mixing), produces 0.5 WER improvement. However, this improvement reduces to nearly 0 when at least 25\% of the data is labelled. 
	
	We hypothesize that the benefits seen with $\mathcal{A} \cup \mathcal{B}$ at \textit{0\%-labelled} come from $\mathcal{A}$ acting as a form of regularisation when most of the client data is pseudo-labelled. Exploring continual learning techniques for this problem is a promising future direction. It is worth noting the surprisingly small difference in WER seen with and without mixing labelled and unlabelled data -- this is likely due to the nature of LS rather than a general observation for ASR datasets (as discussed in \cite{Dimitriadis2020AFA,DBLP:conf/icassp/GaoPZFGBL22}). 
	\vspace{-0.125cm}
	
	% We also compare training with $\mathcal{A} \cup \mathcal{B}$ and only with $\mathcal{B}$. While $\mathcal{A}$ always remains labelled and on the server, $\mathcal{B}$ changes from labelled to pseudo-labelled data, but always remains distributed amongst clients.
	
	% this setup is similar to \cite{DBLP:journals/corr/abs-2107-06877/fedstar} but for the ASR task instead of audio classification. 

	% For Table \ref{tab:sec1b}, rather than selecting whole clients as labelled or unlabelled, we split each client's data into labelled and pseudo-labelled disjoint subsets. We then use this data in its entirety when this client is selected for training. This technique of training is similar to \cite{DBLP:journals/corr/abs-2107-06877/fedstar} but for the ASR task instead of audio classification.

	% We only see a marginal difference -- at most 4.8\%, between WER of models for 0\%-25\% labelled clients.

	% In the experiments, we implemented ASR models using the transformer recipe of SpeechBrain \cite{speechbrain} for LibriSpeech\footnote{\url{https://github.com/speechbrain/speechbrain/blob/develop/recipes/LibriSpeech/ASR/transformer}}. The recipe implements an end-to-end transformer ASR architecture with a Conformer encoder \cite{conformer}. The Conformer configuration follows the Conformer (S) described in \cite{conformer} and we used a transformer language model. The loss is computed using a weighted average of the CTC loss and KL-divergence with label smoothing. The label smoothing parameter is 0.15. 
	
	% TODO: some expt lines taken out we need to put them in a plot

	\subsection{Analyses with Different Labelled Data Regimes}
	We explore the effect of changing the amount of labelled data $\mathcal{A}$ to produce $\text{\supl}_{\text{seed}}$ and further with pseudo-labelled $\mathcal{B}$ with \fnst in Table \ref{tab:sec3}. We used ${\vert \mathcal{A} \vert \in \{120, 240, 360, 480\}}$ hours and $\vert \mathcal{B} \vert = 480 \text{ hours}$ for all experiments. As expected, decreasing available labelled data for pre-training shows higher WER compared to \fnst results in Table \ref{tab:main-results}. Table~\ref{tab:sec3} also points to a clear correlation between the WER of the initial pre-trained model and that of \fnst.
	\vspace{-0.20cm}
	
	\subsection{Analyses with Varying Pseudo-Labelling Frequency}
	\label{sec:ps-every-round}
	We ran an experiment to test our alternative pseudo-labelling strategy in which $\pslbl$ is performed every round $t$ for clients $\mathcal{S}_t$. We provide the results in Table~\ref{tab:pslbl-fly}. Although this strategy performs very similarly to our original approach in terms of WER, the pseudo-labelling step is highly computationally expensive, increasing the overall FL experiment time by 10x. We leave optimizing on this front as future work.
	
	% We found almost no difference in test-clean and limited performance degradation of relative 2.7\% in test-other, making it a viable strategy in realistic FL settings. However due to pseudo-labelling every round being computationally expensive, the overall runtime of the experiment grew by almost 10x. 
	\vspace{-0.25cm}
	
	% sbb28 : 480b_ps-lbl_480a_lbl_pslblonce_1/20220316_1514_45.050045
	% sbb21: 480b_ps-lbl_480a_lbl_onthefly_1/20220216_1531_47.385274
	
	%limited to 50 rounds due to excessive training time.
	\begin{table}[ht!]
		\caption{Comparison of pseudo-labelling strategies, using labelled $\mathcal{A}$ and pseudo-labelled $\mathcal{B}$ for 50 rounds.}
		\vspace{-0.125cm}
		\label{tab:pslbl-fly}
		\centering
		\resizebox{1\columnwidth}{!}{%
			\begin{tabular}{rccc}
				\toprule
				\textbf{Labelling} & \textbf{test-clean} & \textbf{test-other} & 
				\textbf{Wall-clock/FL Round (Mins)} \\
				\midrule
				Once         & 3.31	        & 8.07	        & 1.74         \\
				Every round   & 3.34        & 8.29         & 18.6         \\    \bottomrule
			\end{tabular}
		}
		\vspace{-0.365cm}
	\end{table}

	\vspace{-0.1cm}
	\section{Conclusions}
	
	% \todo[inline]{
	% TODO LIST:
	
	% 4. Insert citations as required \\
	% 5. Update missing results with 1k round expts \\
	
	% }
	
	We have proposed a new method called \fnst for semi-supervised training of ASR models in FL systems. \fnst performs noisy student training to leverage  private unlabelled user data and improves the accuracy of models in low-labelled data regimes using FL. Evaluating \fnst on
	real-world ASR use-cases using the LibriSpeech dataset  with over 1000 simulated FL clients showed
	\textbf{22.5\% relative WERR} over a supervised baseline trained  only with  labelled data available at the server. Our analyses showed that \fnst achieves a WER comparable to fully centralized NST and to supervised training while incurring \textbf{no extra communication overhead} compared to \fedavg. In the future, we plan to employ  \fnst on more challenging datasets, e.g., CommonVoice \cite{DBLP:conf/lrec/ArdilaBDKMHMSTW20}, and to incorporate other methods to learn from unlabelled data, such as Wav2Vec2.0 \cite{DBLP:conf/asru/ChungZHCQPW21}.

	\bibliographystyle{IEEEtran}
	
	\bibliography{mybib}
	
\end{document}